\documentclass[pra,twocolumn,aps,superscriptaddress,showpacs,longbibliography]{revtex4-1}
\usepackage{hyperref}
\hypersetup{
	colorlinks=true,
	citecolor=blue,
	linkcolor=blue,
	urlcolor=blue}
\usepackage{graphicx}
\usepackage[normalem]{ulem}
\usepackage{amsmath}
\usepackage{amsfonts}
\usepackage{amssymb}
\usepackage{epsfig}
\usepackage{mathtools}
\usepackage{braket}
\usepackage[usenames,dvipsnames]{color}
\usepackage{setspace}
\usepackage{bm}
\usepackage{times}
\usepackage{ulem}
\usepackage{dsfont}

\usepackage{xcolor}

\begin{document}
\title{
Quantum phases and spectrum of collective modes in a spin-1 BEC with spin-orbital-angular-momentum coupling}
\author{Paramjeet Banger}\email{2018phz0003@iitrpr.ac.in}
\affiliation{Department of Physics, Indian Institute of Technology Ropar, Rupnagar-140001, Punjab, India}
\author{Rajat}\email{rajat.19phz0009@iitrpr.ac.in} 
\affiliation{Department of Physics, Indian Institute of Technology Ropar, Rupnagar-140001, Punjab, India}
\author{Arko Roy}\email{arko@iitmandi.ac.in}
\affiliation{School of Physical Sciences, Indian Institute of Technology Mandi, Mandi-175075 (H.P.), India}
\author{Sandeep Gautam}\email{sandeep@iitrpr.ac.in}
\affiliation{Department of Physics, Indian Institute of Technology Ropar, Rupnagar-140001, Punjab, India}

\begin{abstract}
Motivated by the recent experiments [Chen {\em et al.}, Phys. Rev. Lett {\bf 121}, 113204 (2018), 
Chen {\em et al.}, Phys. Rev. Lett. {\bf 121}, 250401 (2018)], we investigate the low-lying excitation spectrum of the ground-state
phases of spin-orbital-angular-momentum-coupled (SOAM-coupled) spin-1 condensates. At vanishing detuning, a ferromagnetic SOAM-coupled 
spin-1 BEC can have two ground-state phases, namely coreless and polar-core vortex states, whereas an antiferromagnetic
BEC supports only polar-core vortex solution. The angular momentum per particle, longitudinal magnetization, and excitation frequencies 
display discontinuities across the phase boundary between the coreless vortex and polar-core vortex phases. 
The low-lying excitation spectrum evaluated by solving
the Bogoliubov-de-Gennes equations is marked by avoided crossings and hence the hybridization of the spin and density 
channels. The spectrum is further confirmed by the dynamical evolution of the ground state subjected to a perturbation suitable
to excite a density or a spin mode and a variational analysis for the density-breathing mode.   

\end{abstract}
\maketitle
\section{Introduction}
The experimental realization of spin-orbit (SO) coupling marked an important milestone 
in the field of quantum degenerate Bose gases \cite{lin2011spin,campbell2016magnetic,luo2016tunable}.
The SO coupling in these experiments, coupling the spin and linear momentum of electrically neutral bosons, 
is created by controlling the interaction between atoms and light \cite{lin2011spin,campbell2016magnetic,luo2016tunable,goldman2014light}.
The rich ground-state phase diagram of SO-coupled spin-1 BECs besides having stripe, plane-wave, and 
zero-momentum phases \cite{wang2010spin,martone2016tricriticalities} also admit half-quantum vortex \cite{ramachandhran2012half}, 
vortex-lattice states \cite{ruokokoski2012stationary}, etc. Apart from the ground-state phase diagram, 
collective excitations in trapped SO-coupled BECs, another aspect of fundamental interest \cite{pethick_smith_2008,pitaevskii2016bose}, have been
studied experimentally \cite{PhysRevA.90.063624, PhysRevLett.114.105301} as well as theoretically \cite{PhysRevA.95.033616, PhysRevLett.127.115301} 
in harmonically-trapped SO-coupled pseudospinor BECs. Recently, we studied the collective excitations in 
a quasi-one-dimensional SO-coupled spin-1 BEC with antiferromagnetic interactions at zero and finite 
temperatures \cite{PhysRevA.106.013304}. 

For the last few years, there has been a growing interest in coupling the orbital angular momentum of atoms' center-of-mass 
with their internal spin states using a pair of copropagating Laguerre-Gaussian laser beams with opposite winding numbers.
Commonly known as the spin-orbital-angular-momentum (SOAM) coupling, this feature has been independently demonstrated 
by two experimental groups by coupling two \cite{zhang2019ground} or three magnetic sub-levels of $F = 1$ manifold of $^{87}$Rb atoms
\cite{chen2018spin,chen2018rotating}, thus affirming the validity of earlier theoretical proposals \cite{PhysRevA.92.033615,demarco2015angular,sun2015spin,qu2015quantum,chen2016spin,peng2022spin}. 
There also has been an interest in theoretical models of SOAM coupling inspired 
by Weyl-type SO coupling \cite{xu2020phase}.
In the context of SOAM-coupled pseudospin-1/2 BECs, polarized and zero-momentum phases have been observed
experimentally. Besides these, a stripe, an annular-stripe, a two-vortex molecule, and a vortex-antivortex molecule phases have also
been studied theoretically \cite{qu2015quantum,duan2020symmetry,chen2020angular,chiu2020visible}.  
Theoretical studies on the effects of the ring-trapping potential on the annular-stripe phase in SOAM-coupled pseudospin-1/2 condensate 
have also been carried out \cite{sun2015spin,bidasyuk2022fine}. 
  
Along with studies on equilibrium ground-state phase diagrams, spectroscopic studies have been carried out on SOAM-coupled pseudospin-1/2 BEC 
\cite{chen2020angular,vasic2016excitation,chen2020ground}. In particular, the low-lying excitation
spectrum, including breathing and dipole modes, have been studied for the half-skyrmion and vortex-antivortex phases~\cite{vasic2016excitation,chen2020angular}.
Additionally, the ground-state phases and excitation spectrum have been studied for 
pseudospin-1/2 BEC with higher order SOAM coupling~\cite{chen2020ground}.

 In the experimental realizations of SOAM coupling in the spin-1 spinor BEC of $^{87}$Rb atoms~\cite{chen2018spin,chen2018rotating}, a
Gaussian and a Laguerre-Gaussian beam co-propagating along $z-$direction were considered, leading to an orbital angular-momentum transfer of $\hbar$ to the atoms.
Considering a theoretical SOAM-coupling model with an angular-momentum transfer of $2\hbar$ \cite{chen2016spin}, the ground-state phase diagram and the dynamics
ensuing on sudden quench of quadratic Zeeman terms have been studied. The different considerations of angular-momentum transfer to the atoms yield different
single-particle Hamiltonians and, consequently, different phase diagrams. In this context, considering the experimentally realized
SOAM coupling~\cite{chen2018spin,chen2018rotating}, the detailed phase diagrams and excitation spectrums of SOAM-coupled spin-1 BECs with polar and
ferromagnetic spin-exchange interactions have not been yet theoretically studied. To the best of our knowledge, collective excitations
of an SOAM-coupled spin-1 BEC have not been studied, irrespective of the theoretical models employed.
This sets the stage for the present work.
With inspiration drawn from the experimental research reported in \cite{chen2018rotating} and an aim to bridge 
the research gap, our objective is to study the excitation spectrum of the ground-state phases observed in 
SOAM-coupled spin-1 BECs.
The excitation spectrum, calculated by solving the Bogoliubov-de Gennes (BdG) equations, is supported by the time evolution
of the expectation of the physical observables with an aptly chosen perturbation being added to the Hamiltonian at time $t = 0$.
For the sake of comprehensiveness, we additionally employ the variational method to analytically calculate the frequency of the density-breathing mode.

The paper is organized as follows. In Sec.~\ref{Sec-II}, we present the Hamiltonian 
describing an SOAM-coupled spin-1 BEC in cylindrical coordinates and the reduction 
to a quasi-two-dimensional (quasi-2D) formulation through a set of coupled Gross-Pitaevskii equations (GPEs). In Sec.~\ref{Sec-III},
we discuss the ground-state phases of SOAM-coupled ferromagnetic and polar BECs in the limit
of vanishing detuning. In Sec.~\ref{Sec-IV}A, we discuss the spectrum of the noninteracting 
SOAM-coupled spin-1 BEC, and follow it with the collective excitations of the interacting SOAM-coupled 
spin-1 BECs in \ref{Sec-IV}B.  In Sec.~\ref{Sec-IV}C, we explore the effect of detuning on the ground-state phases and excitation spectrum.
In Sec.~\ref{Sec-IV}D, we study real-time dynamics of the perturbed 
ground state to illustrate the ensuing dynamics in the density and spin channels. In 
Sec.~\ref{Sec-IV}E, the variational method to study a few low-lying modes, is discussed, which
is followed by the summary of key results in Sec.~\ref{concl}.

\section{Model}
\label{Sec-II}
In this work, we consider SOAM-coupled 
spin-1 BECs in which the orbital angular momentum of the center of the mass of the atoms is synthetically coupled to their internal spin 
states~\cite{chen2018spin,zhang2019ground}. In the cylindrical coordinate system, the non-interacting (single-particle) part 
of the Hamiltonian for the spinor BEC is~\cite{chen2018spin,chen2018rotating}
\begin{align}
\label{single_pa}
H_{\rm s}=&\Bigg[-\frac{\hbar^2}{2M}\frac{\partial}{r \partial r}
        \left(r\frac{\partial}{\partial r}\right)+
        \frac{L_z^2}{2M r^2}-\frac{\hbar^2}{2M}
        \frac{\partial^2}{\partial z^2}\\
       & + V({\bf r})\Bigg]\mathds{I} 
+ {\Omega(r) \cos(\phi) S_x-\Omega(r)\sin(\phi)S_y+\delta S_z},\nonumber
\end{align}
where $\mathds{I}$ is a 3$\times$3 identity matrix, $V({\bf r}) = M \omega_0^2 r^2/2+M\omega_z^2 z^2/2$ constitutes the external harmonic 
potential to trap the atoms of mass $M$, $L_z=-i \hbar \partial/\partial\phi$ is the angular momentum operator,
$\Omega(r) = \Omega_0 \sqrt{e}(r/r_0)e^{-r^2/2 r_0^2}$ is the Raman-coupling strength 
with $\Omega_0$ and $r_0$ as the Rabi frequency and the radius of the maximum-intensity (cylindrical) surface~\cite{chen2018spin,chen2018rotating}, respectively,
$\delta$ is the Raman detuning, and $S_x, S_y$ and $S_z$ are irreducible representations of the spin-1 angular momentum 
operators. Under mean-field approximation, the interacting part of the  Hamiltonian $H_{\rm int}$
is given by \cite{kawaguchi2012spinor}
{\begin{equation}
H_{\rm int} = \frac{c_0}{2} \rho+
\frac{c_1}{2} {\bf F}.{\bf S}
\label{int_h}
\end{equation}}
with $c_0$ and $c_1$ as the mean-field interaction parameters.
The total density of the system is given by $\rho$, ${\bf F} = (F_x, F_y, F_z)$ is the spin-density vector{\color{black}, and ${\bf S}=(S_x,S_y,S_z)$}.
Since the SOAM coupling is restricted to the radial plane, 
and we consider $\omega_z \gg \omega_0$, the dominant dynamics is constrained to the same plane with frozen axial degrees of freedom.
We can then integrate out the $z$ degree of freedom from the condensate wave function 
and describe the system as quasi-2D on the radial $r$-$\phi$ plane. 
Starting from the Hamiltonian $H=H_{\rm s} + H_{\rm int}$, in polar coordinates, we obtain the following coupled 
quasi-2D GPEs in dimensionless form
\begin{subequations}
\begin{eqnarray}
i\frac{\partial \psi_{\pm1}}{\partial t} &=&
\mathcal{H}\psi_{\pm1} +{c_1}(\rho_0  \pm \rho_{-})\psi_{\pm1}+{c_1}\psi_{\mp1}^*\psi_0^2  \nonumber \\
&&\pm \delta  \psi_{\pm1} +\frac{\Omega(r)}{\sqrt{2}} e^{\pm i \phi} \psi_0 ,\label{inter_gp2d-1}\\
i\frac{\partial \psi_0}{\partial t} &=&
\mathcal{H}\psi_0 +{c_1}\rho_{+}\psi_0+ 2{c_1}\psi_{+1}\psi_{-1}\psi_0^*
\nonumber\\
 &&+\frac{\Omega(r)}{\sqrt{2}} (e^{-i \phi} \psi_{+1} +e^{i \phi} \psi_{-1}),\label{inter_gp2d-2}
\end{eqnarray}
where
\end{subequations}
\begin{align} 
\mathcal{H} &=-\frac{1}{2}\frac{\partial}{r \partial r}\left(r\frac{\partial}{\partial r}\right)+\frac{L_{\rm z}^2}{2r^2}+ \frac{r^2}{2}+{c_0}\rho,\nonumber\\
\rho &= \sum_{j=\pm 1,0} \rho_j,~\rho_j = |\psi_j|^2,~\rho_{\pm}  = \rho_{+1}\pm\rho_{-1}.\nonumber\\
\end{align}
Under geometric renormalization, in terms of $s$-wave scattering lengths $a_0$ and $a_2$ in the total spin $0$ and $2$ channels, respectively, $c_0$ and $c_1$ take the form
\begin{equation}
c_0 = \sqrt{8\pi\alpha}\frac{N(a_0+2a_2)}{3{a_{\rm osc}}},\quad
c_1 = \sqrt{8\pi\alpha}\frac{N(a_2-a_0)}{3{a_{\rm osc}}} 
\label{interaction2d}
\end{equation}
denoting the spin-independent and spin-dependent interactions, respectively. The anisotropy parameter $\alpha= \omega_z/\omega_0$ is defined 
to be the trapping frequency ratio along the axial to the radial direction, and $N$ is the total number of atoms.  
The units of length, time, energy, and energy eigenfunctions are considered to be 
$a_{\rm osc} = \sqrt{\hbar/(M\omega_0)}$, $\omega_0^{-1}$, $\hbar\omega_0$, and $a_{\rm osc}^{-1}$, 
respectively, and $\int r\rho(r) drd\phi= 1$.
\section{Ground state quantum phases of SOAM coupled spinor BEC}
\label{Sec-III}

To understand the intercomponent phase relationship imposed by various competing terms in
the Hamiltonian, we consider a generic circularly symmetric {\em ansatz}, $\psi_j = f_j(r)e^{i(w_j \phi + \beta_j)}$,
for the component wavefunctions, where $w_j$ and $\beta_j$ are, respectively, the phase-winding number and constant phase associated with
the radially-symmetric real function $f_j$. The phase-dependent part of the interaction energy is minimized, provided~\cite{PhysRevA.66.023602}
\begin{subequations}\label{wn_cond1}
\begin{align}
w_{+1}-2w_0+w_{-1} &= 0,\\ 
\beta_{+1}-2\beta_0+\beta_{-1} &= \begin{cases} 2n\pi\quad{\rm for}~c_1<0,\\
(2n'+1)\pi\quad{\rm for}~c_1>0,
\end{cases}
\end{align}
\end{subequations}
where $n$ and $n'$ are integers.
Similarly, SOAM-part of the energy is minimized if
\begin{subequations}\label{wn_cond2}
\begin{align}
w_{+1} - w_{0} &= 1,\quad w_0-w_{-1} = 1,\\
\beta_{+1} - \beta_0 &= (2p+1)\pi,\quad \beta_0-\beta_{-1} = (2p'+1)\pi,
\end{align}
\end{subequations}
where $p$ and $p'$ are again integers.
If the conditions on the winding numbers in Eq.~(\ref{wn_cond2}a) are satisfied, the
condition in Eq.~(\ref{wn_cond1}a) is satisfied too. On the other hand,
conditions between the constant phase factors in Eqs.~(\ref{wn_cond1}b) and (\ref{wn_cond2}b)
can be simultaneously satisfied for $c_1<0$ only.

To further substantiate the intercomponent phase relationships imposed by SOAM-coupling, we 
extract ${\cal S} = S_x \cos\phi-S_y \sin\phi$ ~\cite{duan2020symmetry} from the single-particle Hamiltonian $H_{\rm s}$.
In the limit when $\Omega_0$ is large, $c_1$ dependent part of the Hamiltonian can be neglected, and the phase structure
of the emergent ground-state solution is mainly determined by ${\cal S}$ via its minimum energy eigen spinor.
The normalized eigen spinor of ${\cal S}$ with minimum eigen energy $-1$ can be written as
$(e^{i(m+1)\phi},~-\sqrt{2}e^{im\phi},~e^{i(m-1)\phi} )^T/2$ with $m$ being any integer.
The phase structure of this eigenspinor is consistent with phase relations in Eqs.~(\ref{wn_cond2}a)-(\ref{wn_cond2}b).
With an increase in $m$, there is an energy cost from the phase-dependent part of the kinetic
energy suggesting that only small values of phase-winding numbers may emerge. Numerical
results confirm this, where we obtain a solution corresponding to $m = 0$ in large $\Omega_0$ limit irrespective
of the nature of spin-exchange interactions.
The spinor part of the ground state in this limit tends to approach the aforementioned eigenstate of ${\cal S}$ with $m = 0$. 

Various numerical techniques have been employed in the literature
to study spinor BECs in quasi-one-dimensional, quasi-two-dimensional, and three-dimensional settings~\cite{roy-2020,kaur2021fortress,ravisankar2021spin, banger2021semi,banger2022fortress}. 
In practice, we choose the finite-difference method and choose different initial guess solutions as an input to Eqs.~(\ref{inter_gp2d-1})-(\ref{inter_gp2d-2})
to arrive at ground-state solutions. As an example, we take initial states $\Psi\sim e^{-r^2/2} \times (e^{i(m+1)\phi},~-
\sqrt{2}e^{im\phi},~e^{i(m-1)\phi} )^T/2$, with different values of $m$. Besides these initial states, we consider a random initial 
guess where $\psi_j(r)$ are complex Gaussian random numbers.

\begin{figure}[ht]
\begin{center}
\includegraphics[width=0.95\columnwidth]{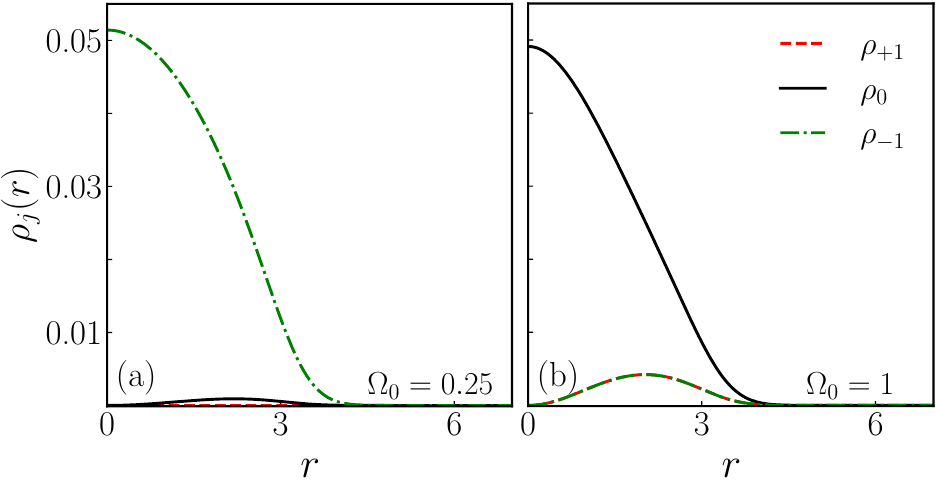}
\caption{(Color online) 
Ground-state densities of the SOAM-coupled $^{87}$Rb spin-1 BEC with
$c_0=121.28$ and $c_1=-0.56$ corresponding to $N = 5000$ for  (a) $\Omega_0 = 0.25$ and (b) 
$\Omega_0 = 1$. The $j = +1,0,$ and $-1$ spin components carry phase winding numbers of $+2,+1,$ and $0$, 
respectively, in (a) and $+1,0,$ and $-1$, respectively, in (b). {As discussed
in the text, the various quantities in this and the rest of the figures are dimensionless.}
}
\label{phase}
 \end{center}
 \end{figure}

At the outset, motivated by the experimental realization of the SOAM-coupled BECs~\cite{chen2018spin,chen2018rotating} 
using spin-1 $^{87}$Rb atoms, we validate our numerical simulations to study and emulate the observed ground-state quantum 
phases of the ferromagnetic system in the absence of detuning $\delta=0$ first and later with $\delta\ne0$. It is to be noted that in the experiments ~\cite{chen2018rotating}, 
both zero and non-zero values of detuning have been considered. 
Similar to the experiment, we consider the 
$^{87}$Rb atoms confined in an anisotropic harmonic trap with $\omega_0 = 2\pi \times 140$ Hz and $r_0=15$ $\mu$m~\cite{chen2018rotating}.
However, we take $\ omega_z = 2\pi \times 2400$ Hz, enabling us to perform quasi-2D simulations. Here 
$a_0=101.8a_B$ and $a_2=101.4a_B$ with $a_B$ as the Bohr radius~\cite{PhysRevLett.88.093201}. The ground-state densities and phase distributions, 
obtained numerically by solving the coupled GPEs (\ref{inter_gp2d-1})-(\ref{inter_gp2d-2}) with imaginary-time propagation, for given $\Omega_0$ and $N$, 
are in qualitative agreement with the experimental results. 
The ground-state densities calculated for a pair of $\Omega_0$ values with $N = 5000$ are shown in Figs.~\ref{phase}(a)-(b).
 \begin{figure}[ht]
 \begin{center}
 \includegraphics[width=0.98\columnwidth]{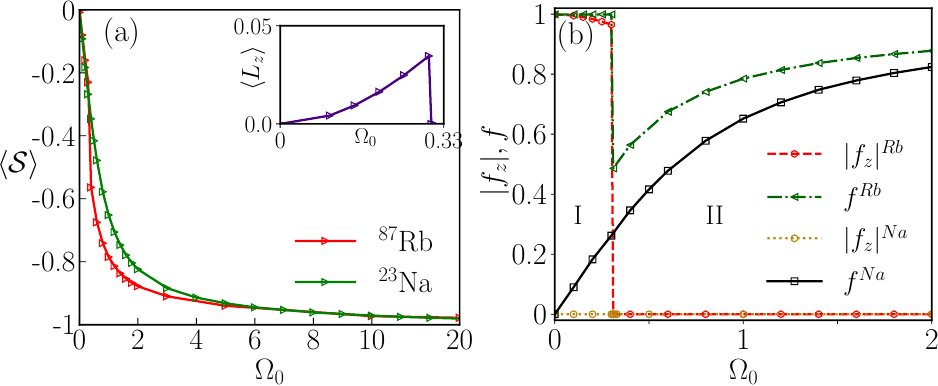}
 \caption{(Color online)
(a) $\left<\cal S\right>$ 
as a function of SOAM-coupling strength $\Omega_0$ for
$^{87}$Rb with $c_0=121.28$ and $c_1=-0.56$ and $^{23}$Na
with $c_0=121.35$ and $c_1=3.80$. Inset in (a): $\langle L_z \rangle$
for $^{87}$Rb as a function of $\Omega_0$.
(b) $|f_z|$ and $f$ for $^{87}$Rb and $^{23}$Na as a function of SO coupling strength $\Omega_0$.
The $c_0$ and $c_1$ for $^{87}$Rb and $^{23}$Na are the same as those in (a). 
}
 \label{sz_phase}
\end{center}
 \end{figure}
\begin{figure}[ht]
\begin{center}
\includegraphics[width=0.99\columnwidth]{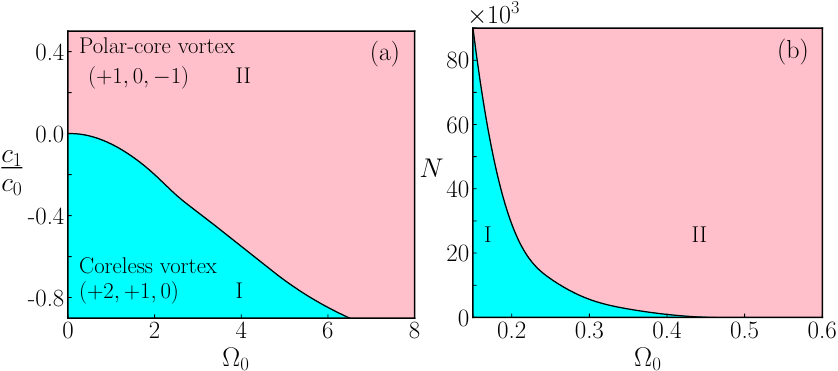}
\caption{(Color online) 
The ground-state phase diagrams in (a) $c_1/c_0$-$\Omega_0$ and
(b) $N$-$\Omega_0$ planes. In (a) $c_0$ was kept fixed at $121.28$ while
varying $c_1$. In (b) $c_1/c_0 = -0.0046$ corresponding to $^{87}$Rb. 
}
\label{phase_diagram}
\end{center}
\end{figure}
For $\Omega_0 = 0.25$, the solutions with $(+2,+1,0)$ and $(0,-1,-2)$ 
phase-winding numbers are two 
degenerate ground states, and with
$\Omega_0 = 1$, 
$(+1,0,-1)$ state is obtained as the ground state solution. 
As we vary $\Omega_0$ from 0 to 20, at small $\Omega_0$, due to the co-action of 
spin-dependent interaction term and SOAM coupling, (+2,+1,0)-type solution appears as the ground state. 
After a critical value of coupling strength (say $\Omega_0^c$), 
$\Omega {\cal S}$ primarily dictates the nature of the solution to result in $(+1,0,-1)$-type
phase. The condition $\langle{\cal S}\rangle\approx-1$ is satisfied in this latter phase
for sufficiently large $\Omega_0$ as shown in Fig.~\ref{sz_phase}(a), which indicates that
no further phase can be expected with higher $\Omega_0$. We term these two phases I and II.
In contrast to $^{87}$Rb, $(+1,0,-1)$-type is the single ground state phase for $^{23}$Na
with $c_1>0$. In this case too, $\langle{\cal S}\rangle\approx-1$ at large $\Omega_0$
as shown in Fig.~\ref{sz_phase}(a).

Longitudinal magnetization per particle 
$f_{z}=\int {F_{z}} d{\bf r}$, spin expectation  per particle $f=\int {|{\bf F}|} d{\bf r}$ where
 $|{\bf F}|={\sqrt{F_x^2+F_y^2+F_z^2}}$, and angular
 momentum per particle $\langle L_z\rangle$ can be used to characterize these ground-state phases. 
In the ferromagnetic domain with $c_0 = 121.28$ and $c_1 = -0.56$, for $\Omega_0<=\Omega_0^c =0.3$, 
i.e. in phase I, $\langle L_z \rangle\ne 0$ and increases continuously
as shown in the inset of Fig. \ref{sz_phase}(a), whereas 
$|f_z|\approx1$ and $f=1$ as shown in Fig.~\ref{sz_phase}(b).
For $\Omega_0 > \Omega_0^c$, the transition to phase II is accompanied by discontinuities 
in $\langle L_z \rangle, |f_z|$, and $f$, whereas the former two reduce to zero, the latter becomes
less than one. In the antiferromagnetic domain, e.g. with $c_0= 121.35$ and $ c_1 = 3.8$, there is no phase transition
with an increase in $\Omega_0$ resulting in smooth behaviour of the same quantities. Here
$f$ asymptotically approaches one, whereas $|f_z|$ and $\langle L_z\rangle$, expectantly, remain zero.
  
\begin{figure}[ht]
\begin{center}
\includegraphics[width=0.48\textwidth]{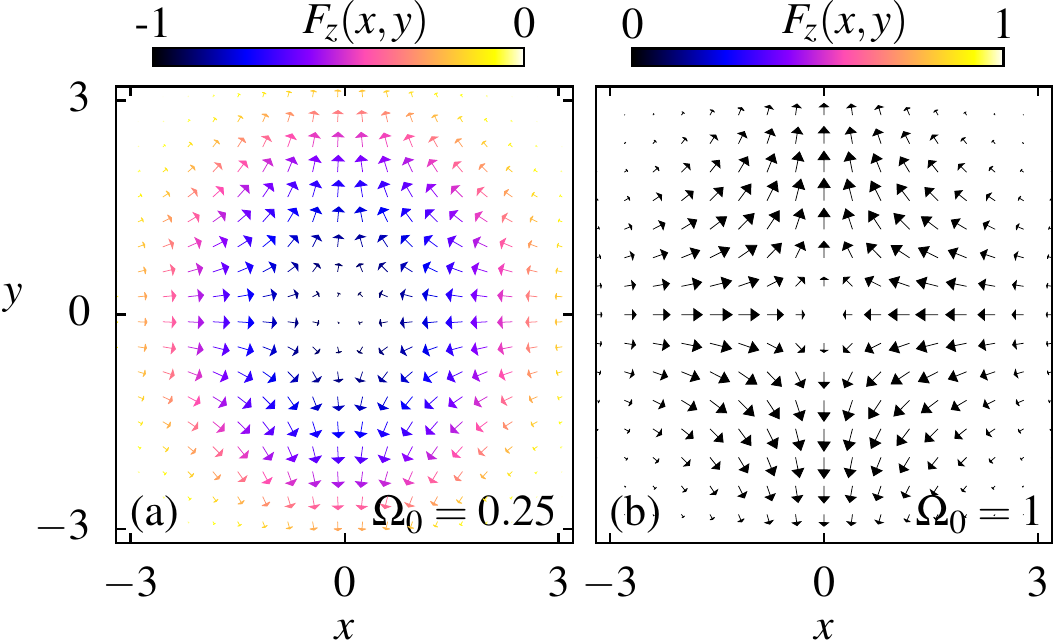}
\caption{(Color online)  (a) and (b) show the spin-texture for 5000 atoms of $^{87}$Rb system 
with the coupling strength $\Omega_0=0.25$ and $\Omega_0=1$, respectively.  The length of the arrows shows the projection of ${\bf F}(x,y)$ on the $x$-$y$ plane, and the color-bar indicates its component along $z$ axis; ${\bf F}(x,y)$ vector field lies on the $x$-$y$ plane in (b).
}
\label{texture}
\end{center}
\end{figure}

Furthermore, we calculate the ground state phase diagrams in
$c_1/c_0$-$\Omega_0$ plane, where we fix $c_0 = 121.28$ and vary $c_1$,
and $N$-$\Omega_0$ plane for fixed $c_1/c_0 = -0.0046$ which corresponds to $^{87}$Rb. 
The ratio $c_1/c_0$ may be manipulated experimentally by tuning one
of the scattering lengths by optical Feshbach resonance \cite{theis2004tuning}. These two
are respectively shown in Figs.~\ref{phase_diagram}(a)-(b), thus again illustrating
that an antiferromagnetic BEC has one ground-state phase in contrast to the ferromagnetic one. 
It can be seen that with a decrease in $c_1$ (keeping $c_0$ fixed) in the ferromagnetic phase, 
the domain of phase I increases, whereas with an increase in the number of atoms 
(keeping $c_1/c_0$ fixed), it decreases. 
Phase I and II also have distinctive topological spin textures {\color{black}  ${\bf F} = (F_x, F_y, F_z)$ }. For the solutions in Figs. \ref{phase}(a) and (b) 
spin-textures are shown in Figs. \ref{texture}(a) and (b), respectively. 
 The spin-textures in Fig.~\ref{texture}(a) and (b) are in agreement with those
reported in \cite{chen2018rotating}; at the centre, ${\bf F}$ points along negative $z$ direction in Fig.~\ref{texture}(a), whereas it is zero in Fig.~\ref{texture}(b).
The details of the spin-textures allow the identification of phases I and II with the coreless vortex and polar-core vortex states, respectively. 
It is to be noted in Ref. \cite{chen2016spin} the two reported circularly symmetric phases correspond to 
$(-4,-2,0)$- and $(-2,0,+2)$-type solutions, which are distinct phases I and II in the present work.

\section{Collective Excitation Spectrum}
\label{Sec-IV}
To study the excitation spectrum, we exploit the innate circular symmetry of the Hamiltonian. To this end, 
we perform a local spin rotation about $\hat{z}$ by the azimuthal angle -$\phi$ to remove the $\phi$-dependence from
the Hamiltonian. As a result, the order parameter $\Psi = (\psi_{+1}, \psi_0, \psi_{-1})^T$ is transformed to 
$e^{-i S_z\phi}{\Psi} =(e^{-i\phi}\psi_{+1}, \psi_0, e^{i\phi} \psi_{-1})^T$, and the transformed Hamiltonian
takes form
\begin{align}
\label{transformed_h}
{H} =& \left[-\frac{1}{2}\frac{\partial}{r \partial r}\left(r\frac{\partial}{\partial r}\right)+\frac{(L_z+S_z)^2}{2 r^2}+ V(r)\right]\mathds{I}\nonumber  \\ 
&+\Omega(r)S_x  + H_{\rm int},
\end{align}
where {${H}_{\rm int} = c_0\rho/2+
c_1{\bf {F}}.{\bf S}/2$}. The Hamiltonian in Eq.~(\ref{transformed_h}) is circularly symmetric, and one can seek
the simultaneous eigen functions of $H$ and $L_z$ with fixed angular momentum $l_z = 0, 1, \ldots$.
For example, the solutions presented in Figs. \ref{phase}(a) and (b) can now be seen as
corresponding to $l_z=1$ and $0$, respectively.  The single-particle Hamiltonian in  Eq. (\ref{single_pa}) is symmetric under the
transformation defined by an operator $ {\cal R} = \exp(-i S_x \pi)K$, where $K$ is complex-conjugation
operator. This implies that for any $l_z\ne 0$, there will be two degenerate solutions connected $\cal R$. For example,
for $l_z = 1$, the degenerate counterpart with $l_z = -1$ corresponds to $(0,-1,-2)$ phase-winding numbers in the component
wavefunctions.

We use the Bogoliubov approach to study the excitation spectrum. 
{
In which we consider the fluctuations to the ground state by writing the perturbed order parameter
as
\begin{equation}\label{pop}
    {\Psi}(r,\phi,t) = e^{-i\mu t +i(l_z+S_z)\phi} [\Psi_{\rm eq}(r) +\delta {\Psi}(r,t)e^{i l_q \phi}],
\end{equation}
where $\Psi_{\rm eq}(r)=(R_{+1}(r), R_{0}(r), R_{-1}(r))^{T}$ is the radial part of the order parameter with  $R_{j}$ as the radial wavefunction
corresponding to the $j^{\rm th}$ spin component, $\mu$ is the chemical potential, and $l_q = 0, \pm1, \pm2,\ldots$ is the magnetic quantum number
associated with the angular momentum of the quasiparticle excitations.}
The details of the Bogoliubov-de Gennes (BdG) analysis are given in the Appendix. 

\subsection{Non-interacting system}
To understand the effect of coupling strength, we first study the single-particle 
excitation spectrum. The ground-state solution has phase-winding numbers $(\pm 1,0)$
in $ j = \pm 1,0$ spin states, respectively.
\begin{figure}[ht]
\begin{center}
\includegraphics[width=0.45\textwidth]{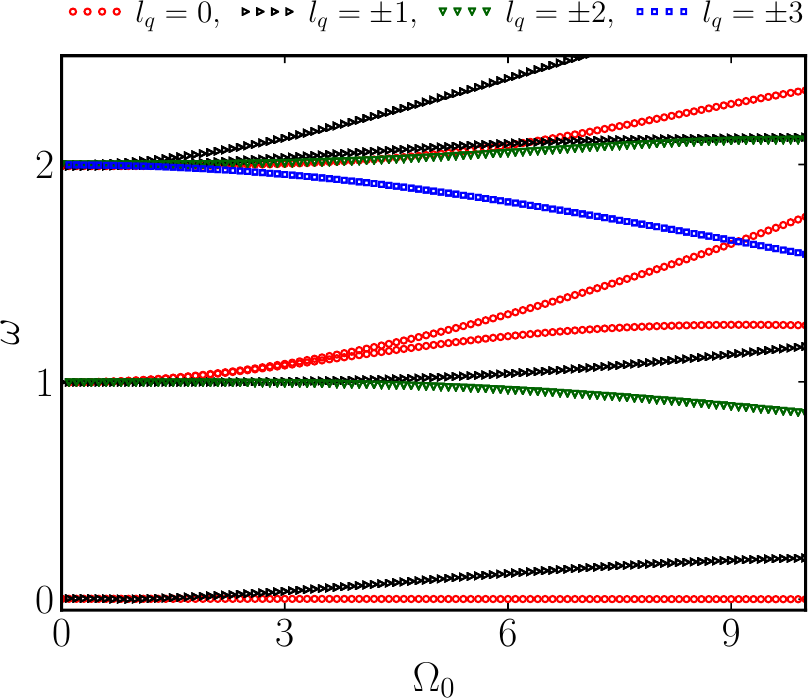}
\caption{(Color online) Single particle excitation spectrum for spin-1 BEC as a function of SOAM coupling strength $\omega_0$.
}
 \label{single_spectrum}
 \end{center}
 \end{figure}
The excitation spectrum is shown in Fig.~\ref{single_spectrum}. For $\Omega_0=0$, the $ n^{\rm th} $  energy level is $3(n+1)$-fold degenerate,
as the single-particle Hamiltonian is identical with a system of three decoupled isotropic two-dimensional harmonic oscillators.
For example, excitations with
energies $0$ and $1$ are three- and six-fold degenerate, respectively.
The SOAM-coupling lifts the degeneracies partially. 
For example, for $\Omega_0\ne 0$, there is only one zero energy excitation; similarly, the red lines 
in the spectrum in Fig.~\ref{single_spectrum} correspond to non-degenerate excitations, 
whereas the black ones to two-fold degenerate modes. The non-degenerate modes have the magnetic quantum number
of the excitation $l_q = 0$, whereas modes with two-fold degeneracy have $l_q \ne 0$.

\subsection{Interacting spin-1 BEC}
Here we study the excitation spectrum (a) as a function of $\Omega_0$ for fixed $c_0$ and $c_1$
and (b) as a function of $N$ for fixed $\Omega_0$ and $c_0/c_1$ ratio. Both $\Omega_0$ and $N$ can
be varied in an experiment \cite{chen2018spin,chen2018rotating}. As was discussed in Sec.~\ref{Sec-III}, for $c_1<0$, 
both phase I and II can appear as the ground-state phases with a variation of either $\Omega_0$ or $N$. We primarily
consider $^{87}$Rb BEC in the following discussion.
\begin{figure}[ht]
\begin{center}
\includegraphics[width=0.45\textwidth]{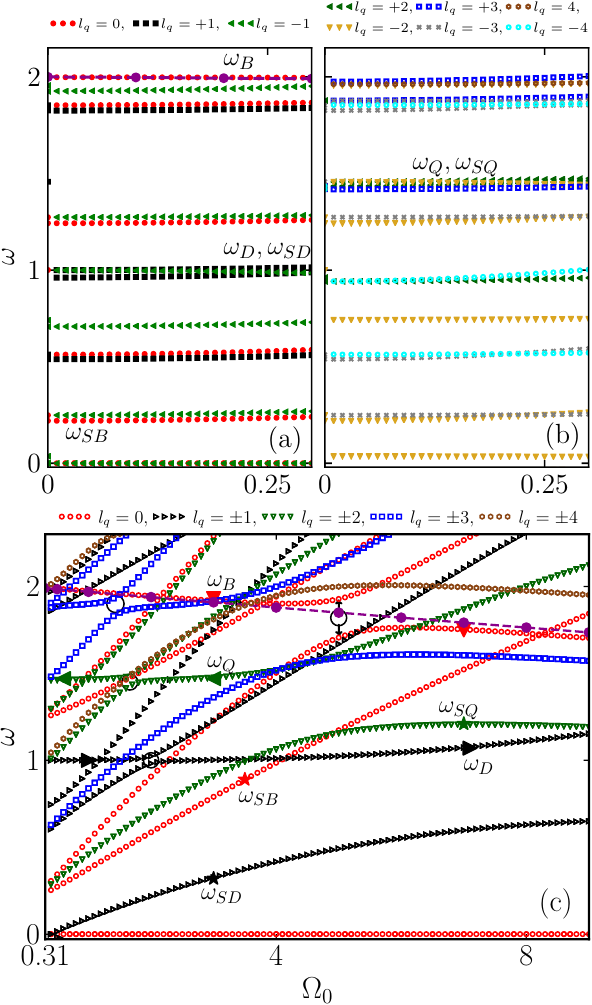}
\caption{(Color online) Low-lying excitation spectrum of $^{87}$Rb SOAM-coupled spin-1 BEC with
$c_0 = 121.18$ and $c_1 = -0.56$ as a function of coupling strength $\Omega_0$ of phase I with $l_q = 0,\pm 1$ in 
(a) and $l_q = \pm2,\pm3, \pm 4,\ldots$ in (b); among the named modes, $l_q = 0$ for density- and spin-breathing, $l_q = +1$ for density-dipole, $l_q = -1$ for spin-dipole,
$l_q = +2$ for density-quadrupole, and $l_q = -2$ for spin-quadrupole modes. (c) shows the same for phase II, where $l_q = 0$ for density- and spin-breathing, $l_q = \pm1$ for density- and spin-dipole,
$l_q = \pm2$ for density- and spin-quadrupole modes. In (a) and (c), the dashed magenta-colored line is the variational
estimate for the density-breathing mode. 
}
\label{inter_spect}
\end{center}
\end{figure}

{\it Phase I:} Here we consider $c_0 = 121.18$ and $c_1 = -0.56$ and vary $\Omega_0$. 
The excitation spectrum for phase I is shown in Fig.~\ref{inter_spect}(a) for $l_q = 0, \pm1$ and (b) for $|l_q|\ge2$. 
The modes with frequencies
$1$ and $2$ are, respectively, dipole and density-breathing modes in Fig.~\ref{inter_spect}(a) . This identification of a mode is based on the real-time evolution of the
expectation of a suitably chosen observable, as will be discussed in the next subsection. The presence of ferromagnetic interactions further
aids the lifting of the degeneracy, in this case between the modes with magnetic quantum numbers $\pm l_q$, which are degenerate at the
single-particle level. We have confirmed this, for example, by examining the excitation spectrum
of a system with $c_0 = 121.18$ and $c_1 = -0.6c_0\ll -0.56$ (not shown here), 
where the non-degenerate nature of the spectrum is clearly seen. 
In phase I, there are two zero-energy Goldstone modes corresponding to two broken continuous symmetries, 
namely gauge and rotational symmetry. The latter corresponds to the symmetry transformation generated by $L_z$.
 
{\it Phase II:}
As already mentioned in Sec. \ref{Sec-III}, the transition from phase I to II occurs at $\Omega_0>0.3$
for $c_0 = 121.18$ and $c_1 = -0.56$. The transition
is accompanied by the discontinuities in the excitation spectrum. The excitation spectrum for phase II is shown in Fig.~\ref{inter_spect}(c).
Here among the low-lying modes are dipole and breathing modes corresponding to both density and spin channels. 
Both density- and spin-dipole modes are doubly degenerate corresponding to magnetic quantum number $l_q = \pm 1$.
On the other hand, both density- and spin-breathing modes are non-degenerate with 
$l_q = 0$. At small values of $\Omega_0$, the energies of the spin modes are less than their
density-mode analogues. There is a single zero-energy mode due to the broken gauge symmetry in this phase.
Besides these modes, the density- and spin-quadrupole modes are also marked
in the excitation spectrum in Figs.~\ref{inter_spect}(a)-(c).  As the collective excitations characterize a system’s response to small perturbations, these can be experimentally studied by using Bragg spectroscopy~\cite{PhysRevA.90.063624, PhysRevLett.114.105301}.

\begin{figure}[h]
\begin{center}
\includegraphics[width=0.46\textwidth]{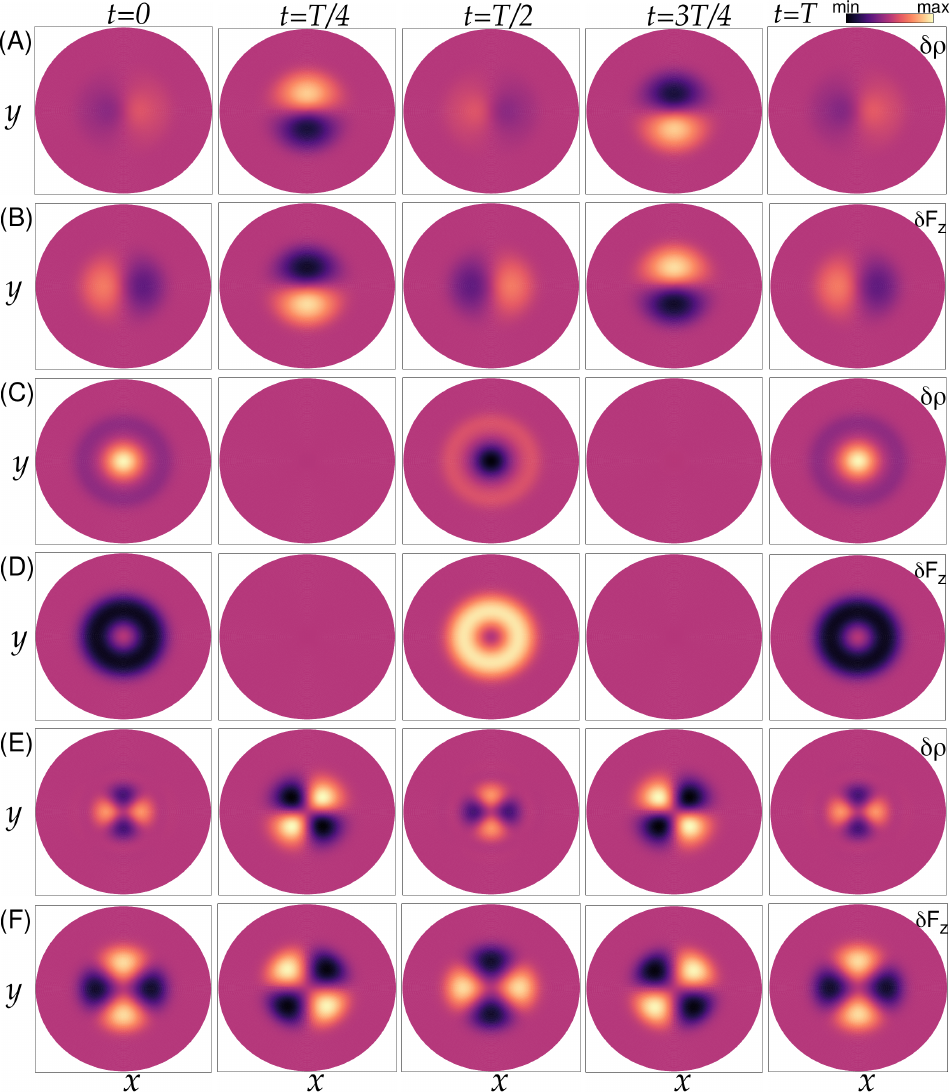}
\caption{(Color online) Density and longitudinal magnetization density fluctuations 
at $t = 0, T/4, T/2, 3T/4,$ and $T$ where $T = 2\pi/\omega$ with $\omega$ as the mode-frequency:  
(A) $\delta \rho(x,y,t)$ for the density-dipole mode with $\omega_{\rm D} = 1$, (B) $\delta F_z(x,y,t)$ for the spin-dipole mode with
$\omega_{\rm SD} = 0.08$, (C) $\delta \rho(x,y,t)$ for
the density-breathing mode with $\omega_{\rm B} = 1.97$, (D) $\delta F_z(x,y,t)$ for the spin-breathing modes with $\omega_{\rm SB} = 0.37$, 
(E) $\delta \rho(x,y,t)$ for the density-quadrupole
mode with $\omega_{\rm Q} = 1.46$, and (F) $\delta F_z(x,y,t)$ for the spin-quadrupole mode with $\omega_{\rm SQ} = 0.46$. The
box size in each subfigure is $6.4a_{\rm osc}\times6.4a_{\rm aosc}$. 
}
\label{fluctuation1T}
\end{center}
\end{figure}

Additionally, the variation in SOAM-coupling strength leads to avoided crossings between the pairs of excitations,
a few of which are identified by the black circles in Fig.~\ref{inter_spect}(c). We observe that the avoided crossing
occur between the density and spin oscillations associated with the same magnetic quantum number $l_q$. In the vicinity of the avoided crossing, the roles of 
the density and spin modes are interchanged as shown in Fig.~\ref{inter_spect}(c).  We study this mode-mixing by examining the density $(\delta \rho)$  and spin fluctuations $(\delta F_x, \delta F_y, \delta F_z)$ yielded by the perturbed order parameter and defined as 
\begin{subequations}
\begin{align}
\delta \rho=&2 {\rm Re} \sum_j \psi_j \delta \psi_j^*,\\
\delta F_x =& \sqrt{2} {\rm Re}(\psi_{+1}\delta \psi_0^*+
\psi_0\delta \psi_{+1}^* 
+\psi_{-1}\delta \psi_0^* +\nonumber\\&
\psi_0 \delta \psi_{-1}^*),\\
\delta F_y =& -\sqrt{2} {\rm Im}( -\psi_{+1} \delta \psi_0^*+
\psi_0\delta \psi_{+1}^* 
+\psi_{-1}\delta \psi_0^* -\nonumber\\&
\psi_0\delta \psi_{-1}^*) ,\\
\delta F_z =& 2{\rm Re}(\psi_{+1} \delta \psi_{+1}^*-
\psi_{-1}\delta \psi_{-1}^*),
\end{align}
\end{subequations}
where '${\rm Re}$' and '$\rm Im$' denote the real and imaginary part, respectively. For a pure density mode, one would expect that $\delta \rho \ne 0$
and  $\delta F_{\nu} = 0$, similarly for a pure spin mode one would
expect that $\delta \rho = 0$ and at least one of the $\delta F_{\nu} \ne 0$. The order-parameter fluctuation $\delta \Psi(r,\phi,t)$, and hence
density and spin fluctuations,
can be constructed with the Bogoliubov quasiparticle amplitudes $u$ and $v$ corresponding to the frequency $\omega$ of the mode {\color{black} as  
$\delta \psi_j(r,\phi,t) \propto e^{i(l_z+j+l_q)\phi}\left[u_j(r) e^{-i\omega t}-v_j^*(r) e^{i\omega t}\right]$.}
\begin{figure}[h]
\begin{center}
\includegraphics[width=0.44\textwidth]{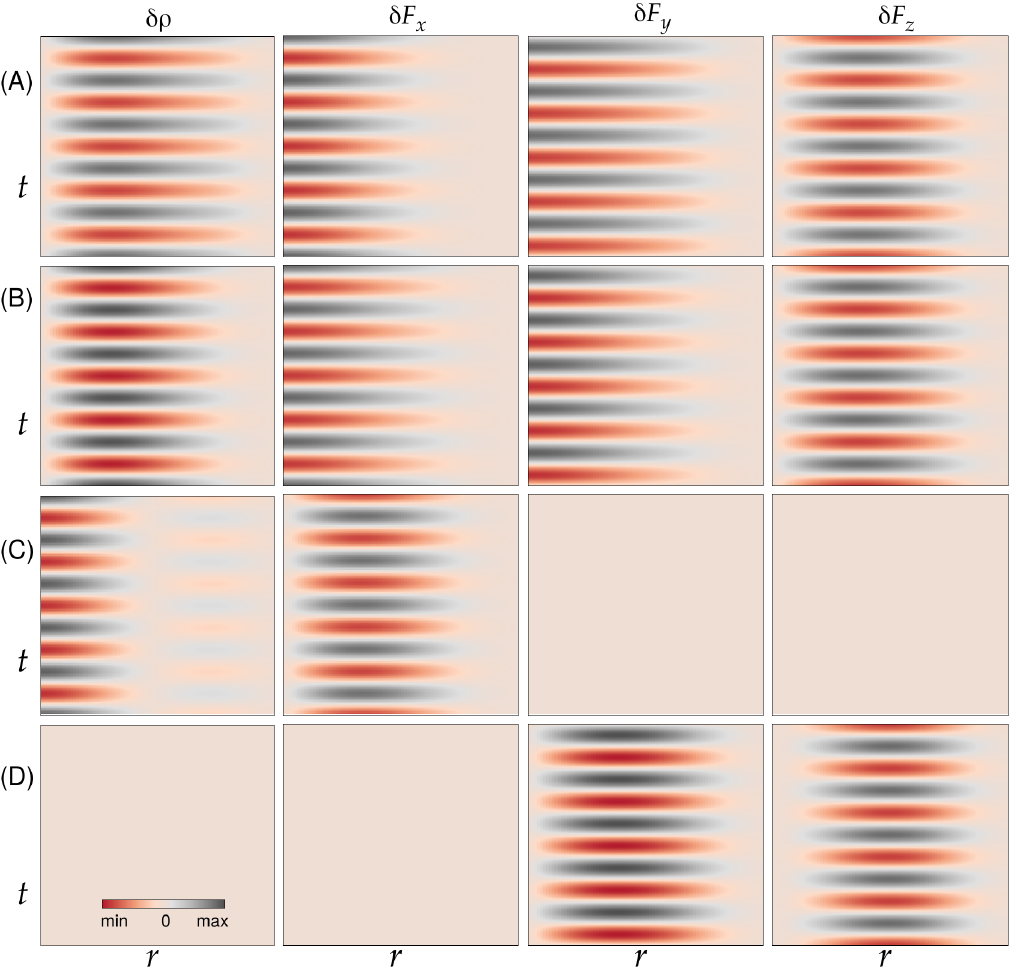}
\caption{(Color online)
(A) shows the density fluctuations, $\delta \rho(r,\phi=0,t)$,
and spin-density fluctuations, $\delta F_{\nu}(r,\phi=0,t)$, with $\nu =x,y,z$ corresponding to $\omega_{\rm D}=1$. (B)-{\color{blue}(D)} present the same for $\omega_{\rm SD}=0.08$, $\omega_{\rm B}=1.97$, $\omega_{\rm SB}=0.37$, respectively.  
The radial and time extents, along horizontal and vertical directions, respectively, in each subfigure 
are $4 a_{\rm osc}$  and $5 T$, respectively, where $T = 2\pi/\omega$ is the time period of the
corresponding mode with $\omega$ frequency. 
{The presence of both density and spin fluctuations in (A)-(C) is an
outcome of the avoided crossing between the pairs of modes in the excitation spectrum shown in Fig.~\ref{inter_spect}(c).}
}
\label{fluctutation}
\end{center}
\end{figure}
In the excitation spectrum in Fig.~\ref{inter_spect}(c) at $\Omega_0=1$, the density- and spin-dipole modes'
frequencies are $\omega_{\rm D}=1$ and $\omega_{\rm SD}=0.08$, respectively, and the density- and spin-breathing modes' frequencies 
are $\omega_{\rm B}=1.97$ and $\omega_{\rm SB}=0.37$, respectively. One can see that the density-dipole, density-breathing, 
and spin-dipole modes encounter
avoided crossings, whereas the spin-breathing mode does not. This observation agrees with the density and spin-density fluctuations evaluated along the $\phi = 0$ line and shown in Figs.~\ref{fluctutation}(A)-{\color{blue}(D)}. 
For the density-dipole mode with $\omega_{\rm D}=1$, both density and spin channels are excited as is seen
from  $\delta \rho(r,\phi =0,t)$ and $\delta F_\nu(r,\phi =0, t)$ in Fig.~\ref{fluctutation}(A), where $\nu = x,y,z$. Similarly, number density, longitudinal $(F_z)$, 
and transverse magnetization ($F_x, F_y$) densities oscillate in time, corresponding to the spin-dipole mode in Fig. \ref{fluctutation}(B), and density-breathing 
mode ends up exciting both the number and transverse magnetization densities in Fig. \ref{fluctutation}(C). On the other hand, the spin-breathing mode 
excites the spin channel alone in Fig. \ref{fluctutation}(D). The density- and spin-quadrupole modes 
too excite both the density and spin fluctuations which are not shown.
This mode mixing indicated by both density and spin fluctuations is absent in 
quasi-one-dimensional SO-coupled BECs where any collective excitation yields either density or spin fluctuations \cite{PhysRevA.106.013304}.
The nomenclature of the modes
in Figs. \ref{inter_spect}(a)-(c) is consistent with the density, $\delta\rho(x,y,t)$, and longitudinal magnetization
density, $\delta F_z(x,y,t)$, fluctuations corresponding to density, breathing, and quadrupole modes in 
Fig. \ref{fluctuation1T} shown at $t=0, T/4, T/2, 3T/4$, and $T$ instants, where $T$ is the period of the collective 
excitation.


\begin{figure*}[ht]
\begin{center}
\includegraphics[width=0.95\textwidth]{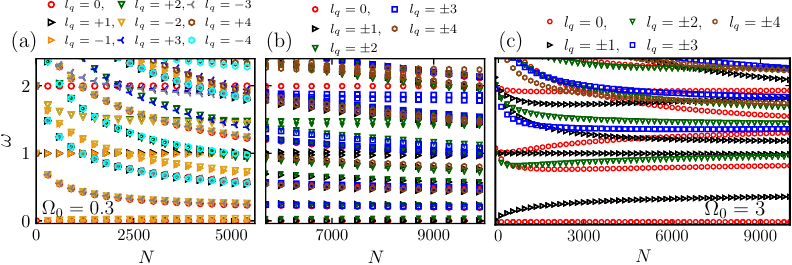}
\caption{(Color online) Low-lying excitation spectrum for $^{87}$Rb spin-1 BEC with $c_1/c_0= -0.0046$ 
as a function of the number of atoms $N$: (a)-(b) for $\Omega_0=0.3$ with a phase transition from phase I
to II at 
$N$ = 5700 and (c) $\Omega_0=3$. (a) corresponds to the spectrum of phase I, whereas (b) and (c) correspond
to the spectrum of phase II. The different point styles in (a)
signify non-degenerate modes with different $l_q$, while in (b) and (c), 
'red circle', {'black right-pointing triangle, 'green down-pointing triangle'}, and 'blue square' correspond, respectively
to the modes with $l_q = 0, \pm 1, \pm 2$, $\pm 3$, and  so on. 
}
\label{natom_rb}
\end{center}
\end{figure*}

Next, we study the excitation spectrum as a function of $N$ for $c_1/c_0 = -0.0046$.
Here first, we fix $\Omega_0$ to $0.3$, where a phase transition from phase I to II
occurs at $N = 5700$. The excitation spectrum, in this case, for phase I and II
are shown in Fig.~\ref{natom_rb}(a) and (b).
The same for $\Omega_0 = 3$ is shown in Fig.~\ref{natom_rb}(c), where phase II is
the ground state phase with no phase transition. The modes in phase II are, again, either non-degenerate
or with two-fold degeneracy. For SOAM-coupled $^{23}$Na BEC with $c_0 = 121.35$ and $c_1=3.8$
the excitation spectrum, which is not shown here, is similar to the spectrum in Fig.~\ref{natom_rb}(c) with
some quantitative differences attributable to different $c_1$ values. 


{\subsection{Non-zero detuning}
\begin{figure}[h]
\begin{center}
\includegraphics[width=0.3\textwidth]{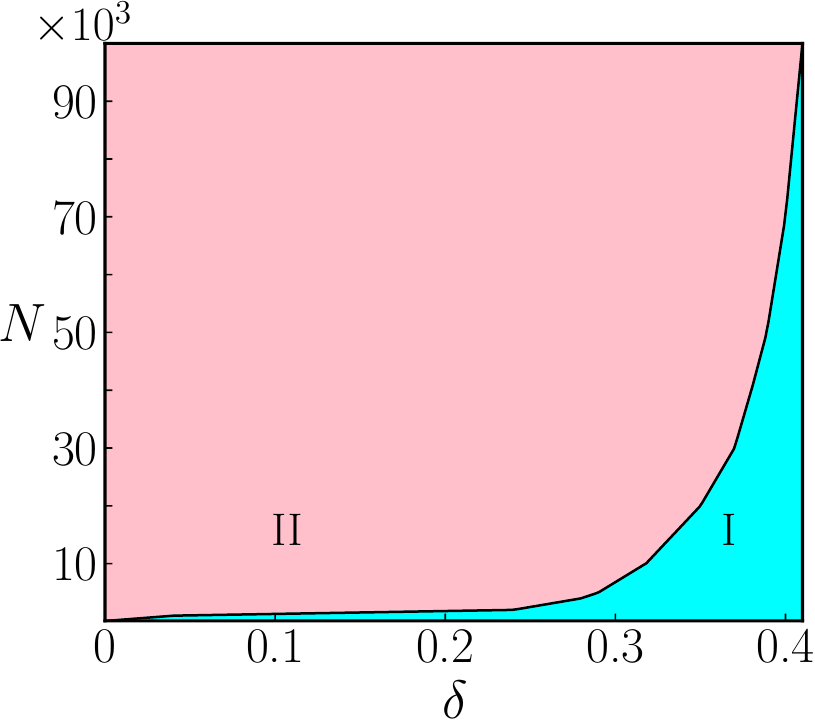}
\caption{(Color online) The ground-state phase diagrams in $N$-$\delta$ plane for $c_1/c_0 =-0.0046$ 
corresponding to $^{87}$Rb spin-1 BEC and $\Omega_0=5$. 
}
\label{delta+ph}
\end{center}
\end{figure}
In this subsection, we consider the effects of the detuning on the phase diagram and excitation
spectrum. In Fig.~\ref{delta+ph}, we show the phase diagram in the number of atoms versus
the detuning ($N$-$\delta$) plane for a constant coupling strength of $\Omega_0 = 5$ and $c_1/c_0=-0.0046$ corresponding
to $^{87}$Rb. We observe for a small value of $\delta$, the polar-core vortex (phase II) emerges
as the ground-state solution. However, at a critical detuning $\delta_c$, 
a phase transition from (+1,0,-1)-type solution (phase II) to (+2,+1,0)-type solution (phase I) occurs.  For example, for
$\Omega_0 = 5$ and $N = 5000$ corresponding to $c_0 = 121.18$, $c_1 = -0.56$,  
the phase transition occurs at $\delta_c=0.3$. Phase II at smaller detuning and phase I at
larger detuning values in Fig.~\ref{delta+ph} is in qualitative agreement with the experimental findings \cite{chen2018rotating}.
It is worth noting that the presence of $\delta$ in the Hamiltonian leads to the breakdown of the symmetry defined
by $\cal R$. As a result, (+2,+1,0) and (0,-1,-2)-type solutions corresponding to $l_z = 1$ and $-1$, respectively, are
no longer degenerate. To illustrate the effect of detuning on the excitation spectrum, we contrast the
collective excitation spectrum of the condensate with $N = 5000$, $\Omega_0 = 5$ for (a) $\delta = 0$ and (b) $\delta = 0.2$.
The ground-state phase in both these cases is phase II as can be seen from the phase diagram in Fig.~\ref{delta+ph}.
The excitation frequencies as a function of $l_q$ for these two cases are shown in Figs. \ref{detun_spec} (a)-(b).
As discussed in the Appendix, the presence of detuning leads to the lifting of the degeneracies in the excitation spectrum about 
$l_q=0$. The low-lying modes have been identified in Fig. \ref{detun_spec} (a)-(b). In Fig. \ref{detun_spec}(a), the density-dipole, spin-dipole, 
density-quadrupole, and spin-quadrupole exhibit two-fold degeneracies corresponding to $\pm l_q$. However, in the presence of $\delta$, 
all these modes become non-degenerate and are highlighted in \ref{detun_spec}(b).}

\begin{figure}[!ht]
\begin{center}
\includegraphics[width=0.45\textwidth]{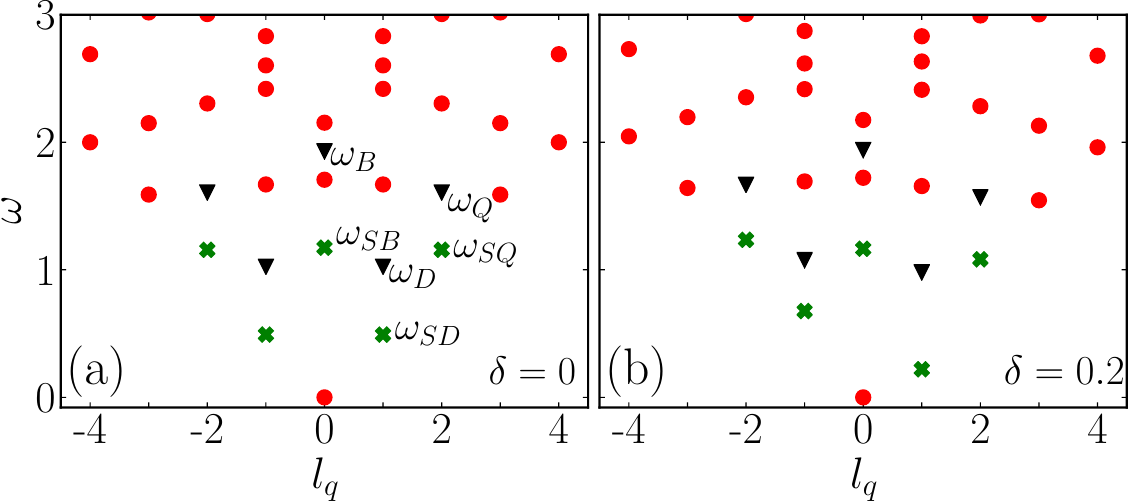}
\caption{{(Color online) Collective excitation spectrum for ferromagnetic $^{87}$Rb spin-1 BEC with 
interaction parameters $c_0=121.28$, $c_1=-0.56$, and coupling stength $\Omega_0=5$ for (a) $\delta=0$ and (b) $\delta=0.2$. 
The 'black {down-pointing} triangles' and 'green crosses' denote the density and spin modes, respectively.}
 }
\label{detun_spec}
\end{center}
\end{figure}
\subsection{Dynamics}
We examine the nature of low-lying collective excitations through the time evolution of the expectation of physical observables, which
also serves to validate our calculation of the excitation spectrum from BdG equations. Here, we consider the Hamiltonian with an appropriately chosen
time-independent perturbation, say $H_{\rm s}'$ added to its single-particle part $H_{\rm s}$. This modifies the coupled GP Eqs. (\ref{inter_gp2d-1})-(\ref{inter_gp2d-2})
with an added term corresponding to $H_{\rm s}'\Psi(r,\phi,t)$ in each equation. We then solve these resultant GPEs over a
finite period of time by considering previously obtained ground-state solutions as the initial solutions at $t = 0$. Numerically, one needs to consider
a two-dimensional spatial grid over here, for which we choose the Cartesian $x$-$y$ grid. 

\begin{figure}[ht]
\begin{center}
\includegraphics[width=0.46\textwidth]{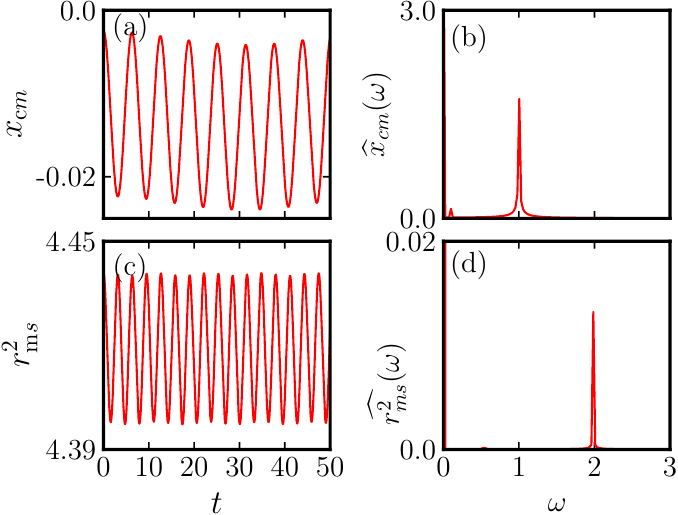}
\caption{(Color online) (a) shows the centre of mass oscillations, i.e. $x_{\rm cm}(t)$ as a function of time and (b) corresponding Fourier transform 
with a dominant peak at $\omega=1$ for $^{87}$Rb spin-1 BEC with
$c_0=121.28$, $c_1=-0.56$, and $\Omega_0=1$. (c) shows the oscillations in the mean square size of the system  $r^{2}_{\rm ms}(t)$ and (d)
the corresponding Fourier transform with a dominant peak at $\omega=1.99$ for the same interaction and coupling strengths. }
\label{DB_mode}
\end{center}
\end{figure}
\begin{figure}[ht]
\begin{center}
\includegraphics[width=0.46\textwidth]{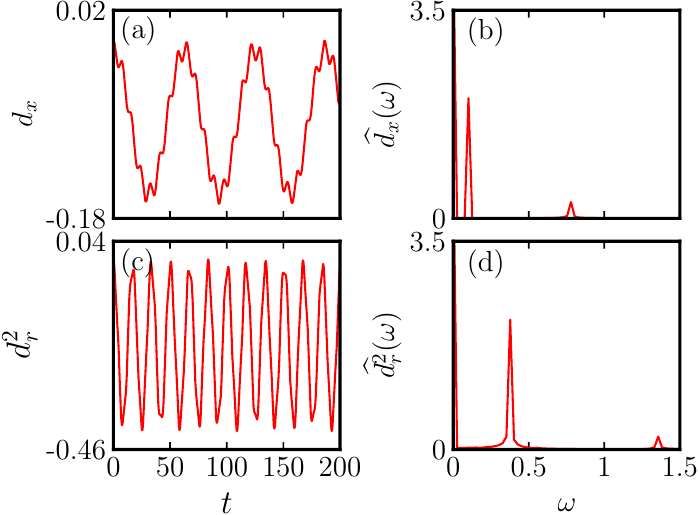}

\caption{(Color online) (a) shows $d_{x}(t)$ as
a function of time and (b) corresponding Fourier transform with a dominant peak at 
$\omega=0.1$ for $^{87}$Rb spin-1 BEC with
$c_0=121.28$, $c_1=-0.56$, and $\Omega_0=1$. Similarly, (c) and (d) show the
$d^{2}_{x}(t)$ and its Fourier transform with 
a dominant peak at $\omega=0.37$ for the same interaction and coupling strengths.}
\label{SDSB_mode}
\end{center}
\end{figure}

We consider $c_0 = 121.28$, $c_1 = -0.56$, and $\Omega_0 = 1$, 
which yielded the ground-state phase in Fig. \ref{phase}(b), as an example set of parameters to study the dynamics.
To excite the density-dipole mode, we take the perturbation $H_{\rm s}' = \lambda x$, where $\lambda\ll 1$. 
We then examine the dynamics of the center of mass of the BEC via $x_{\rm cm}(t) = \langle x \rangle = 
\sum_{j=\pm 1,0}\int x \rho_j(x,y,t) dxdy$ which is plotted in Fig. \ref{DB_mode}(a). 
We also compute its Fourier transform $\widehat{x}_{\rm cm}(\omega)$  to demonstrate that the dominant 
frequency resonates at $\omega=1$ as can be seen in Fig. \ref{DB_mode}(b) and matches
with $\omega_{\rm D} = 1$ in the BdG spectrum in Fig. \ref{inter_spect}(c). We could have
chosen $H_s' = \lambda y$ and then calculated $y_{\rm cm}(t)$ giving us the same excitation frequency.
This is a consequence of the two-fold degeneracy in the density-dipole mode.
We have checked that this mode can also be excited by shifting the minima of the external trapping potential.
This particular way of exciting this mode has direct relevance from an experimental point of view, where
the minima of potential can be easily shifted.
Similarly, to examine the excitation of the density-breathing mode with $H_{\rm s}' = \lambda (x^2+y^2)$, where 
the relevant observable is $r^2 = x^2+y^2$, we calculate mean square radius $r_{\rm ms}^2(t) = \langle r^2\rangle$ 
as a function of time, which is plotted in Fig. \ref{DB_mode}(c). 
The Fourier transform $\widehat{r_{\rm ms}^2}(\omega)$ of $r_{\rm ms}^2(t)$ reveals a dominant peak at 
$\omega=1.99$ in Fig.~\ref{DB_mode}(d) which is close to BdG result of $\Omega_{\rm B} = 1.97$. This mode, again, can be excited by perturbing the trap strength, which can be achieved in an 
experiment with ease and thus giving access to this mode. 
Similarly, the spin-dipole mode can be excited by adding a perturbation $H' = \lambda x S_z$ or $\lambda y S_z$  
with $x S_{z}$ or $y S_{z}$ as the pertinent observable  corresponding to the spin-dipole mode. The
two possible observables again reflect the two-fold degeneracy of spin-dipole modes.  
The time-variation of $d_x(t)= \langle x S_z\rangle = 
\sum_{j= +1,-1} \int x \rho_j(x,y,t) dx dy$
is shown in Fig. \ref{SDSB_mode}(a) and its Fourier transform in Fig. \ref{SDSB_mode}(b) 
has a dominant  peak 
at $\omega=0.1$, which corresponds to the spin-dipole mode labeled in 
Fig. \ref{inter_spect}(c) with $\omega_{\rm SD} = 0.08$. 
Similarly, the spin-breathing mode corresponds to observable $r^2 S_{z}$. 
In Figs. \ref{SDSB_mode}(c) and (d), we show the dynamics of $d_r^2(t) = \langle r^2S_z\rangle$,
i.e. the relative difference in the mean-square radii of the $j = \pm 1$ components and the 
associated Fourier transform, respectively, with a dominant peak at $\omega=0.37$, in agreement with $\omega_{\rm SB}$ 
in Fig. \ref{inter_spect}(c). The very small secondary peaks present in Fig. \ref{DB_mode}(b) and \ref{DB_mode}(d) correspond to the spin-dipole and spin-breathing modes, respectively. These peaks become prominent when subjected to
appropriate perturbations and are observed through relevant observables, as shown in Fig. \ref{SDSB_mode}.
Likewise, the small peaks appear in Fig. \ref{SDSB_mode}(b) and 
\ref{SDSB_mode}(d) also signify modes present in the BdG spectrum.
Finally, the density- and spin-quadrupole modes' frequencies
calculated from the time evolution of $\langle x y\rangle$ and $\langle x y S_{z}\rangle$ are
in agreement with the numbers in Fig. \ref{inter_spect}(c).

\subsection{Variational analysis}
For an SOAM-coupled spin-1 system, a few low-lying modes can be studied using a time-dependent variational method \cite{PhysRevLett.77.5320}. 
For example, to calculate the density-breathing mode in
the absence of detuning, we consider the following variational ansatz
\begin{multline}\label{var_ansatz}
\Psi =\frac{r}{2\sqrt{\pi}\sigma(t)^2} \exp\left[-\frac{r^2}{2 \sigma (t)^2}+i \alpha(t) r^2\right]\times
\begin{pmatrix}
  e^{i (m+1) \phi} \\
  -\sqrt{2} e^{im \phi}\\
 e^{i (m-1) \phi},
\end{pmatrix}
\end{multline}
where $\sigma(t)$ and $\alpha(t)$ are time-dependent variational parameters used to denote the width of condensate and chirp of Gaussian pulse, respectively,  and $m=\pm1$ for phase I or $0$ for phase II. 
The Lagrangian of the system is given by
\begin{equation}
\label{lagran}
 L = \sum_{j} \int dr d\phi \frac{i}{2}\left( \psi_j^*\frac{\partial \psi_j}{\partial t} -\psi_j\frac{\partial \psi_j^*}{\partial t}\right) -E,
\end{equation}
where energy $E$ is defined as 
\begin{align}
E=&\iint
\Bigg[\sum_j\psi_j^*\left\{-\frac{1}{2r}\frac{\partial}{\partial r}
 \left(r\frac{\partial}{\partial r}\right)+
\frac{L_{\rm z}^2}{2r^2}+
\frac{r^2}{2}\right\}\psi_j \nonumber\\ 
&+\frac{c_0}{2}\rho^2 + \frac{c_1}{2}(\rho_{+1} +\rho_0-
\rho_{-1})\rho_{+1}\nonumber\\&+ \frac{c_1}{2}(\rho_{+1} +\rho_{-1}) 
\rho_{0}\nonumber +   \frac{c_1}{2}(\rho_{-1} +\rho_0-
\rho_{+1})\rho_{-1}\nonumber\\&+ {\sqrt{2}}{\Omega(r)} {\rm Re}
(\psi_{+1}^*e^{i\phi}\psi_0 +\psi_{-1}^*e^{-i\phi}\psi_0 )
  \nonumber\\ 
& +2 c_1{\rm Re}(\psi_{-1}^*\psi_0^2\psi_{+1}^*)\Bigg] dr d\phi.
\end{align}
For $ m = \pm1$, the (coupled) Euler-Lagrange equations are
\begin{subequations}\label{EL_eqs}
\begin{align}
\ddot{\sigma(t)} =&\frac{\sigma}{2}\left(\frac{6 \sqrt{2 \pi } \sqrt{e} \Omega_0  \sqrt{\frac{1}{r_0^2}+\frac{2}{\sigma^2}} \left(r_0^7-2 r_0^5 \sigma^2\right)}{\left(2r_0^2+\sigma^2\right)^4}-2\right) \nonumber\\& +\frac{c_0+c_1+10 \pi }{8 \pi  \sigma^3},\\
\alpha=&\frac{\dot{\sigma}}{2\sigma},
\end{align}
\end{subequations}
where $\Dot{}$ denotes the time derivative.
The equilibrium width $\sigma_{0}$ of the condensate satisfies
\begin{equation}
 \frac{c_0+c_1+10 \pi }{4 \pi  \sigma^4_{0}}+\frac{6 \sqrt{2 \pi } \sqrt{e} \Omega_0  \sqrt{\frac{1}{r_0^2}
 +\frac{2}{\sigma^2_{0}}} \left(r_0^7-2 r_0^5 \sigma^2_{0}\right)}{\left(2r_0^2+\sigma^2_{0}\right)^4}=2.\nonumber
\end{equation}
The frequency of the oscillation in width calculated by linearizing Eq.~(\ref{EL_eqs}a) about equilibrium width $\sigma_{0}$ is
\begin{align}
\omega_{\rm B}^{\rm I}=& \left[ \frac{15\sqrt{2\pi }r_0^4\sqrt{e}\sigma_{0} \Omega_0 (3 r_0^2-2\sigma^2_{0}) \sqrt{{2r_0^2}+{\sigma^2_{0}}}+1}{(2 r_0^2+\sigma^2_{0} )^5} +\right. \nonumber\\ & \left.\frac{3(c_0+c_1+10\pi)}{8\pi\sigma^4_{0}}\right]^{1/2}.
\end{align}
Similarly, for $m = 0$ in Eq.~(\ref{var_ansatz}), the density breathing mode is 
\begin{align}
\omega_{\rm B}^{\rm II}=&\left[\frac{15\sqrt{2\pi }r_0^4\sqrt{e}\sigma_{0} \Omega_0 (3 r_0^2-2\sigma^2_{0}) \sqrt{2{r_0^2}+{\sigma^2_{0}}}+1}{(2 r_0^2+\sigma^2_{0})^5}
+\right. \nonumber \\& \left.\frac{3(c_0+c_1+6\pi)}{8\pi\sigma^4_{0}}\right]^{1/2}.
\end{align}
The variationally calculated density-breathing mode's frequency agrees with the values in the BdG spectrum as
demonstrated in Figs.~\ref{inter_spect}(a) and (c) for phases I and II, respectively. As mentioned in
Sec. \ref{Sec-IV}D, density breathing mode can be easily excited by modulating the trapping potential strength
in an experiment.

\section{Summary and conclusions}
\label{concl}
We have investigated the low-lying collective excitations of the coreless vortex and the polar-core vortex phases supported
by the spin-1 BECs with SOAM coupling. The existence of the two phases is seen in the full phase diagrams in the {\em ratio of interaction strengths} versus 
{\em coupling strength} and also the {\em number of atoms} versus {\em coupling strength} planes. We have studied the excitation spectrum as a function of
two experimentally controllable parameters, namely coupling 
strength and the number of atoms. The excitation spectrums are characterized
by the discontinuities across the phase boundary between the two phases and within a phase by avoided crossings between the modes with the same magnetic
quantum number of excitations. The avoided crossings signal the hybridization of the density and spin channels; the nature of spin and density fluctuations 
has indeed confirmed this. Among the low-lying modes, we identify dipole, breathing, and quadrupole modes for density and spin channels.
The frequencies of these named modes are further validated from the time evolution of the expectations of the physical observables when
an apt time-independent perturbation is added to the system's Hamiltonian. An analytic estimate for the density-breathing modes
has also been obtained using the variational analysis. Our results can serve as a benchmark to compute the finite-temperature phase diagram and spin dynamics.
With the experimental observation of collective excitation, dispersion (excitation energies as a function of wavenumber) in Raman-induced
SO-coupled BECs~\cite{PhysRevA.90.063624, PhysRevLett.114.105301}, we expect that our results can also be verified in future SOAM-coupled experiments.
With the advent of box trapping potential~\cite{PhysRevLett.110.200406}, an interesting future direction could be to study an SOAM-coupled BEC in 
such a trap with no rotational symmetry.  



\section*{Acknowledgments}
AR acknowledges the support of the Science and Engineering Research Board (SERB), Department of Science and Technology, Government of India under the project SRG/2022/000057 and IIT Mandi seed-grant funds under the project IITM/SG/AR/87. AR acknowledges National Supercomputing Mission (NSM) for providing computing resources of PARAM Himalaya at IIT Mandi, which is implemented by C-DAC and supported by the Ministry of Electronics and Information Technology (MeitY) and Department of Science and Technology (DST), Government of India. S.G. acknowledges support from the Science and Engineering Research Board, Department
of Science and Technology, Government of India through Project No. CRG/2021/002597.

\begin{widetext}
\section*{Appendix}
{\bf Bogoliubov-de Gennes (BdG) analysis:}
The fluctuation $\delta {\Psi(r,t)}$ to the equilibrium order parameter in Eqn.~(\ref{pop}) is 
$\delta {\Psi}(r,t) = u(r) e^{-i\omega t}-v^*(r) e^{i\omega t}$, where $u(r)$ and $v(r)$ are
Bogoliubov amplitudes and $\omega$ is the excitation frequency. Linearization of the three coupled Gross-Pitaevskii 
Eqs.~(\ref{inter_gp2d-1})-(\ref{inter_gp2d-2}) and the conjugate set of equations using perturbed order parameter 
in Eq.~(\ref{pop}) yields following six-coupled BdG equations:
 \begin{subequations}
 \begin{eqnarray}
\omega u_{+1}   &=&
 \left[-\frac{\nabla^2_r}{2} +\frac{r^2}{2}+ \delta+ \frac{(l_q+l_z+1)^2}{2}-\mu +c_0 (2R_{+1}^2 +R_0^2+R_{-1}^2)+  c_1(2R_{+1}^2+R_0^2-R_{-1}^2)\right]u_{+1} \nonumber \\
&&
 +\left[\frac{\Omega(r)}{\sqrt{2}}+R_{+1} R_0(c_0+c_1) +2 c_1 R_0 R_{-1}\right]u_0 + R_{+1}^2 (c_0+c_1)v_{+1} + R_{+1} R_0(c_0+c_1) v_0\nonumber \\
&& 
 + R_{+1} R_{-1}(c_0-c_1)u_{-1} +(R_{+1} R_{-1} (c_0-c_1)+2c_1R_0^2)v_{-1}, 
\label{bdg-1u}\\
-\omega v_{+1}   &=&
 \left[-\frac{\nabla^2_r}{2} +\frac{r^2}{2}+ \delta+ \frac{(l_q+l_z+1)^2}{2}-\mu +c_0 (2R_{+1}^2 +R_0^2+R_{-1}^2)-  c_1(2R_{+1}^2+R_0^2-R_{-1}^2)\right]v_{+1}\nonumber\\ && + \left[\frac{\Omega(r)}{\sqrt{2}}+R_{+1} R_0(c_0+c_1)
 +2 c_1 R_0 R_{-1}\right]v_0 + R_{+1}^2 (c_0+c_1)u_{+1} + R_{+1} R_0(c_0+c_1) u_0
\nonumber\\ &&
+ R_{+1} R_{-1}(c_0-c_1)v_{-1} +\big[R_{+1} R_{-1} (c_0-c_1)+2c_1R_0^2 \big]u_{-1},
 \label{bdg-1v} \\
  \omega u_0   &=&
\left[-\frac{\nabla^2_r}{2} +\frac{r^2}{2}+  \frac{(l_q+l_z)^2}{2}-\mu +c_0(R_{+1}^2 +2R_0^2+R_{-1}^2)-c_1(R_{+1}^2+R_{-1}^2)\right]u_0 +\left[\frac{\Omega(r)}{\sqrt{2}} -R_{+1}^2(c_0+c_1)\right]u_{+1} \nonumber \\
  &&
  + R_{+1}R_0(c_0+c_1)v_{+1}+(c_0R_0^2+2c_1R_{+1}R_{-1})v_0+\left[\frac{\Omega(r)}{\sqrt{2}}+R_0R_{-1}(c_0+c_1) -2c_2R_{+1}R_0 \right]u_{-1}\nonumber \\
  &&
  + R_{+1}R_{-1}(c_0+c_1)v_{-1}\label{bdg-2u} \\
-\omega v_0  &=&
 \left[-\frac{\nabla^2_r}{2} +\frac{r^2}{2}+  \frac{(l_q+l_z)^2}{2}-\mu +c_0(R_{+1}^2 +2R_0^2+R_{-1}^2)-c_1(R_{+1}^2+R_{-1}^2)\right]v_{0} +\left[\frac{\Omega(r)}{\sqrt{2}} -R_{+1}^2(c_0+c_1)\right]v_{+1} \nonumber \\
  && + R_{+1}R_0(c_0+c_1)u_{+1}+(c_0R_0^2+2c_1R_{+1}R_{-1})u_{0}+\left[\frac{\Omega(r)}{\sqrt{2}}+R_0R_{-1}(c_0+c_1)-2c_1R_{+1}R_0 \right]v_{-1}  \nonumber \\ && + R_{+1}R_{-1}(c_0+c_1)u_{-1}\label{bdg-2v} \\
\omega u_{-1}   &=&
 \left[-\frac{\nabla^2_r}{2} +\frac{r^2}{2}- \delta+ \frac{(l_q+l_z-1)^2}{2}-\mu+c_0 (R_{+1}^2 +R_0^2+2R_{-1}^2)+ c_1(2R_{-1}^2+R_0^2-R_{+1}^2)\right]u_{-1} \nonumber \\&&+\left[\frac{\Omega(r)}{\sqrt{2}} 
+R_0R_{-1}(c_0+c_1)+2c_1R_{+1}R_0\right]u_0 + (c_0-c_1)R_{+1}R_{-1} u_{+1} \nonumber \\&& + (R_{+1}R_{-1}(c_0-c_1)+2c_1R_{0}^2)v_{+1} +R_{-1}^2(c_0+c_1)v_{-1}+ R_{+1} R_0(c_0+c_1) v_0
  \label{bdg-3u} \\
-\omega v_{-1}   &=&
 \big[-\frac{\nabla^2_r}{2} +\frac{r^2}{2}- \delta+ \frac{(l_q+l_z-1)^2}{2}-\mu+c_0 (R_{+1}^2 +R_0^2+2R_{-1}^2)+ c_1(2R_{-1}^2+R_0^2-R_{+1}^2)\big]v_{-1} \nonumber\\&& 
+\big[\frac{\Omega(r)}{\sqrt{2}} 
 +R_0R_{-1}(c_0+c_1)+ 2c_1R_{+1}R_0\big]v_{0} + (c_0-c_1)R_{+1}R_{-1} v_{+1}\nonumber\\&&
 + (R_{+1}R_{-1}(c_0-c_1)+2c_1R_0^2)u_{+1} +R_{-1}^2(c_0+c_1)u_{-1}+ R_{+1} R_0(c_0+c_1) u_0
  \label{bdg-3v} 
 \end{eqnarray}
\end{subequations}
where $\nabla^2_r = -\partial^2/(2\partial r^2)-{\partial}/(2 r\partial r)$, $l_z = 1$ for phase I and $0$ for phase II. 
To solve coupled Eqs.~(\ref{bdg-1u})-(\ref{bdg-3v}), we use
the finite-difference method to discretize these equations over the spatial radial grid \cite{GAO2020109058}, thus transforming the
BdG equations to a matrix eigenvalue equation, which can be solved using 
standard matrix diagonalization subroutines. To discretize the BdG equations, we used a radial grid consisting of 
$N_r = 256$ points with a radial step-size of $\Delta r = 0.05$, which results in a $6N_r\times 6N_r$ matrix eigen-value problem.
It is to noted that Eqs.~(\ref{bdg-1u})-(\ref{bdg-3v}) for $l_z = 0$ and $\delta = 0$, remain invariant if $l_q\ne 0$ is changed to $-l_q$ with simultaneous interchange of the $j = +1$ and $j = -1$ components. It implies that for non-zero magnetic quantum number of excitation ($l_q\ne 0$), 
$\pm l_q$ excitation modes in the single-particle excitation spectrum, viz. Fig.~\ref{single_spectrum}, will be degenerate.
For the same reason, $\pm l_q$ modes with $\l_q\ne 0$ of a polar-core vortex solution are also degenerate, for example in Fig.~\ref{inter_spect}(c).
In the presence of detuning $(\delta \ne 0)$ or angular momentum ($l_z\ne 0$), this invariance is not there, and as a result, $\pm l_q$ excitations with $l_q\ne 0$
are no longer degenerate in the coreless vortex phase.

\end{widetext}
\bibliography{bib_file_new}{}
\bibliographystyle{apsrev4-1}
\end{document}